\documentclass[prl,aps,twocolumn,preprintnumbers,showkeys,superscriptaddress]{revtex4}
\usepackage{epsfig,amssymb,amsfonts,verbatim}

\def\bfl{\begin{flushleft}}
\def\efl{\end{flushleft}}
\def\bfr{\begin{flushright}}
\def\efr{\end{flushright}}
\def\bc{\begin{center}}
\def\ec{\end{center}}
\def\be{\begin{equation}}
\def\ee{\end{equation}}
\def\ba{\begin{eqnarray}}
\def\ea{\end{eqnarray}}
\def\baa#1{\begin{array}{#1}}
\def\eaa{\end{array}}
\def\bw{\begin{widetext}}
\def\ew{\end{widetext}}

\def\text#1{\mbox{#1}}

\usepackage{color}
\usepackage[english]{babel}

\def\bfl{\begin{flushleft}}
\def\efl{\end{flushleft}}
\def\bfr{\begin{flushright}}
\def\efr{\end{flushright}}
\def\bc{\begin{center}}
\def\ec{\end{center}}
\def\be{\begin{equation}}
\def\ee{\end{equation}}
\def\ba{\begin{eqnarray}}
\def\ea{\end{eqnarray}}
\def\baa#1{\begin{array}{#1}}
\def\eaa{\end{array}}
\def\bw{\begin{widetext}}
\def\ew{\end{widetext}}

\def\text#1{\mbox{#1}}

\begin{document}

\title{An alternative normal state $c$-axis resistivity model for high-$T_c$ superconductors}

\author{Andrew Das Arulsamy}
\email{sadwerdna@hotmail.com}

\affiliation{Condensed Matter Group, Division of Exotic Matter,
D423, Puteri Court, No. 1, Jalan 28, Taman Putri, 68000 Ampang,
Selangor DE, Malaysia}

\date{19 November 1999; accepted: 11 January 2001}




\begin{abstract}
An alternative model for $c$-axis resistivity in layered
high-$T_c$ crystalline superconductors is proposed and has been
characterized as an essentially two-dimensional Fermi liquid.
Average ionization energy is included as additional parameter that
determines the concentration of tunnelling electrons between
Cu-O$_2$ layers. This model agrees well quantitatively with the
Bi2212 and Y123 single crystals, and qualitatively with the pure
1212 phase polycrystals.
\end{abstract}


\maketitle



\subsection{1. Introduction}

The out-of-plane conduction mechanism in the normal state of
high-$T_c$ superconductors is intriguing. In the early stages,
facts were pointed out how anomalous various normal state
properties were in the simple framework of Fermi
liquid~\cite{kitazawa1}. This Fermi liquid predicted large Fermi
surface whereas small Fermi surface was expected from the strong
correlation models where the normal state transport properties
favored the latter. Further studies revealed that the conventional
field theoretic many-body treatment does not seem to lead to the
non-Fermi liquid behavior either~\cite{fukuyama2}. Apart from
that, according to RVB theory, the normal state of high-$T_c$
superconductors is all metals in the Cu-O$_2$ planes and is
semiconductor-like for conduction between those
planes~\cite{anderson3}. In addition, Anderson and Zou suggested
that both the $a$-$b$ plane and $c$-axis resistivities can be
fitted very accurately by $\rho = A/T + BT$. It is also believed
that there is a gap induced by the localization of carriers along
$c$ direction, which acts as an anomalous constraint of $c$-axis
hopping rate~\cite{zheng4}. Parallel to this, the $c$-axis normal
state resistivity of high-$T_c$ superconductors were derived via
Fermi-Dirac statistics assuming the only object that were allowed
to tunnel through the Cu-O$_2$ layers are the real electrons as
required by the RVB theory. This phenomenological model is also
used to fit the $a$-$b$ plane resistivity. Furthermore, the
estimation of resistivity transition such as from metallic to
semiconductor or vice versa upon substitution in the pure 1212
phase polycrstals is also reported.

\subsection{2. The model}

The tunnelling electron's distribution can be derived using
Fermi-Dirac statistics and ionization energy as an anomalous
constraint. This derivation involves two restrictive conditions:
i) the total number of electrons in a given system is constant and
ii) the total energy of $n$ electrons in that system is also
constant. Both conditions are as given below

\begin {eqnarray}
\sum_i dn_i = 0.
\end {eqnarray}

\begin {eqnarray}
\sum_i (E_{electron})_i dn_i = 0.
\end {eqnarray}

In condition (2) however, the ionization energy, $E_I$ will be
included as an additional constraint where $E_{electron} =
E_{initial~state} + E_I$. This is to justify that an electron to
occupy a higher state N from the initial state M is more probable
than from state L if the condition $E_I(M) < E_I(L)$ at certain
temperature, $T$ is satisfied. Thus, condition (2) is rewritten as

\begin {eqnarray}
\sum_i (E_{initial~state} + E_I)_i dn_i = 0.
\end {eqnarray}

The importance of this inclusion is that it can be interpreted as
a gap and also, particularly the $E_I$ can be used to estimate the
resistivity transition upon substitution of different valence
state ions. Finally, the probability of an electron to occupy an
energy level $E$ is given by

\begin {eqnarray}
f_e \approx e^{-\mu-\lambda(E + E_I)},
\end {eqnarray}

if $e^{\mu+\lambda(E + E_I)} \gg 1$. Note that the distribution
(4) can be reduced to the standard Fermi-Dirac distribution if
$E_I \rightarrow 0$. The concentration of electrons is given by

\begin {eqnarray}
n = \int_0^{\infty} f_e(E) N_e(E) dE,
\end {eqnarray}

where $\mu = -E_F/kT$, $\lambda = 1/kT$ and $N_e(E)$ is the two
dimensional density of states that was derived from the
schrodinger equation. The solution of integral (5) gives the
concentration of tunnelling electrons between Cu-O$_2$ layers for
high-$T_c$ superconductor as

\begin {eqnarray}
n = \frac{m_e^* kT}{\pi \hbar^2} \exp \left[\frac{E_F -
E_I}{kT}\right].
\end {eqnarray}

This concentration is expressed in term of ionization energy and
$k$ is the Boltzmann constant. $m_e^*$ is the effective electron
mass and $E_F$ is the Fermi level. It is well known that the
scattering among Fermions gives $1/\tau \propto T^2$ and $\rho
\propto T^2$ from the empirical rule~\cite{maeno5,maeno6,maeno7}.
According to the elementary metal theory, the resistivity of metal
is given by $\rho = m_e^*/ne^2\tau$ where $e$ and $\tau$ are the
electron's charge and the mean scattering free time,
respectively~\cite{cyrot8}. Therefore, the phenomenological model
of $c$-axis normal state resistivity can be written in the form

\begin {eqnarray}
\rho_c(T,E_I) = A\frac{\pi \hbar^2}{e^2 k}T \exp \left[\frac{E_I -
E_F}{kT}\right],
\end {eqnarray}

Where $A$ is independent of temperature. This model is comparable
with the form given in Ref.~\cite{zheng4} where in the exponential
term, the gap, $\delta$ is used instead of the term $(E_I - E_F)$
as an anomalous constraint of $c$-axis hopping rate between
Cu-O$_2$ layers. It shows that this exponential term has a similar
meaning to this gap.

\subsection{3. Discussion}

The experimental $\rho_c$ plots of doping dependent Bi2212 and
Y123 single crystals were best fitted quantitatively with the
phenomenological model (7). The Bi2212 samples, 1a and 1b were
annealed at 10 and 10$^{-3}$ atm respectively to control the
oxygen content~\cite{zheng4}. In contrast, all the Y123 samples
have the same composition, which is YBa$_2$Cu$_3$O$_7$, except for
the volume of the unit cell which are 630 $\times$ 640 $\times$ 75
and 390 $\times$ 400 $\times$ 25 $\mu$m$^3$ for samples B and C
respectively~\cite{hagen9}.

The experimental plots and the calculated curves of
$\rho_c(T,E_I)$ versus $T$ for Bi2212 are shown in Fig. 1a and b.
Both the experimental plot and the calculated curve in Fig. 1a can
be separated into two regions: (i) $E_I - E_F$ (110K) $<$ $kT$
(metallic-like behavior at $T$ $>$ 110K) and (ii) $E_I - E_F$
(110K) $>$ $kT$ (semiconductor-like behavior at $T$ $<$ 110K).
That is, $\rho_c(T,E_I)$ property changes from metallic-like
behavior at high temperatures to semiconductor behavior at low
temperatures. At high temperatures, $\rho_c \propto \rho_{ab}$,
suggesting that scattering or fluctuations in the Cu-O$_2$ planes
may dominate the linear term of $\rho_c$~\cite{anderson3,zheng4}.
In addition, the thermal energy, $kT$ is higher than the
constraint, $E_I - E_F$ above 110K which give rise to the
conductivity of tunnelling electrons. At low temperatures however,
$\rho_c$ is incoherent and can be interpreted as an indication of
a low thermal energy as $T$ goes below 110K.

Similarly, the experimental and the calculated resistivity curves
for Y123 samples B and C as shown in Fig. 2a also indicates the
same characteristics, where the metallic-like behavior occurs when
$E_I - E_F$ (155.7K) $<$ $kT$ for sample B and $E_I - E_F$
(139.8K) $<$ $kT$ for sample C. The resistivity for both B and C
increases with temperature at $T$ $<$ 155.7K and $T$ $<$ 139.8K
respectively suggesting the semiconductor-like region starts below
those temperatures. The calculated $\rho_{ab}$ from model (7) for
samples B and C as shown in Fig. 2b were also in good agreement
with the experimental data. However, the $E_I - E_F$ parameter has
a much lower value as expected compared to $c$-axis which can be
due to the low spin fluctuations in the Cu-O$-2$ planes. There is
no significant differences in $A$ and $E_I - E_F$ parameters
between sample B and C for both the $a-b$ plane and $c$-axis
although the ratio $a/c$ differs in large magnitude, which are
approximately 8 for B and 16 for C.

The resistivity transition from sample 1a to 1b as the O$_2$
content decreases in Bi$_2$Sr$_2$CaCu$_2$O$_y$ is due to the
transition of $E_I - E_F$ (110K) $<$ $kT_r$ to $E_I - E_F$ (390K)
$>$ $kT_r$, $T_r$ is the room temperature. Therefore, the
concentration of electrons, $n$ between layers in sample 1b is
small due to insufficient thermal energy since $kT_r <$ 390K.
Table 1 lists $A$ and $E_I - E_F$ parameters in detail for the
calculated curves. In the $\rho(T)$ transition estimation,
Tl-Sr-(Y$_{1-f}$X$_f$)-Cu-O system is considered for an example
where the average $E_I$ for X$^{z+}$ and Y$^{A+}$ ions in that
system can be calculated using

\begin {eqnarray}
E_I[X^{z+}] = \sum_i^z \frac{E_{Ii}}{z}
\end {eqnarray}

and

\begin {eqnarray}
E_I[Y^{A+}] = \sum_i^A \frac{E_{Ii}}{A}.
\end {eqnarray}

$i$ = 1, 2,...z and $i$ = 1, 2,...A. $z^+$ and $A^+$ is the
valence state of X and Y respectively. If $E_I$[X$^{z+}$] $<$
$E_I$[Y$^{A+}$] then the substitution of Y with X will reduce $E_I
- E_F$ parameter thus, reduces $\rho_c(T)$ as the curves in Fig. 3
shows. Furthermore, the $\rho_c(T)$ rises rapidly with $E_I - E_F$
for 100K compared to 300K due to the low tunnelling electron's
concentration. This figure also reveals that the room temperature
$c$-axis resistivity increases exponentially with the systematic
substitution of different valence state ions. Table 2 summaries
the estimated $\rho(T)$ for single and double substituted
polycrystalline high-$T_c$ of pure 1212 phase superconductors.
Substitution of Sr$^{2+}$, Er$^{3+}$, Pr$^{3+}$ and Y$^{3+}$ with
Ba$^{2+}$, Sr$^{2+}$, Sr$^{2+}$ and Ca$^{2+}$ respectively reduces
$\rho(T)$ since the ionization energies of the latter is lower
than the former. On the other hand, the substitution of Sr$^{2+}$
with Tm$^{2+}$, Sm$^{2+}$, Eu$^{>2+}$ and Dy$^{>2+}$ increases
$E_I$ hence, $\rho(T)$, therefore the metallic to
semiconductor-like transition can be expected and vice versa for
the former. Here, $\rho_c(T)$ is assumed to be roughly $\rho(T)$
due to polycrystallinity of the samples.

\subsection{4. Conclusions}

Ionization energy has been used in the Fermi-Dirac statistics as
an additional restrictive condition. The calculation for the
tunnelling electron's concentration was done by employing the
two-dimensional Fermion characteristics. The proposed model (7)
agrees well quantitatively with the experimental $\rho_c(T)$ of
Bi2212 and Y123 single crystals and also predicts the resistivity
will increase with the average ionization energy. In addition,
both the experimental data of $\rho_c(T)$ and $\rho_{ab}(T)$ can
be fitted with the proposed model. Besides that, the estimated
$\rho(T)$ transition upon substitution of ions with larger or
smaller ionization energies also agrees well with the pure 1212
phase polycrystals.

\begin{figure}
\caption {Temperature dependence of the $c$-axis resistivities of
Bi2212 single crystals annealed at a) 10 and b) 10$^{-3}$ atm.}
\label{fig1}
\end{figure}

\begin{figure}
\caption {Experimental (dotted) and calculated (solid line) plots
for YBa$_2$Cu$_2$O$_7$ single crystals. a) $c$-axis and b) $a-b$
plane resistivities.} \label{fig2}
\end{figure}

\begin{figure}
\caption {The variation of ($E_I-E_F$) parameter with $c$-axis
resistivity at 100 and 300K.} \label{fig3}
\end{figure}

\begin{table}
\caption {Detailed values of temperature independent $A$ and
$E_I-E_F$ parameters.} \label{Table:I}
\end{table}

\begin{table}
\caption {Average $E_I$, estimated and experimental, $\rho(T)$ of
pure 1212 phase polycrystalline high-$T_c$ superconductors.}
\label{Table:II}
\end{table}

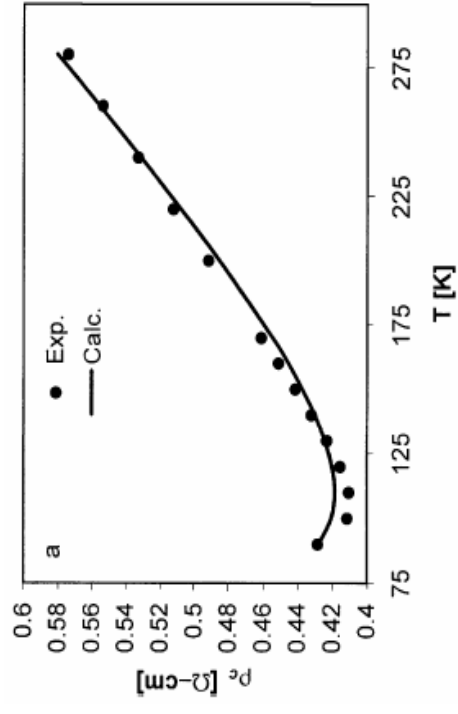
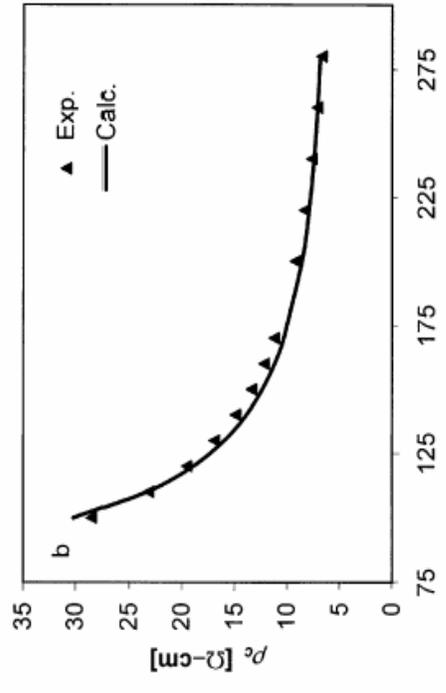
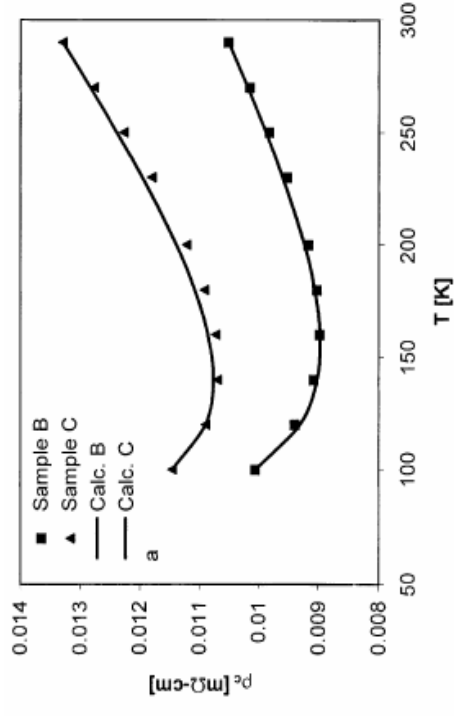
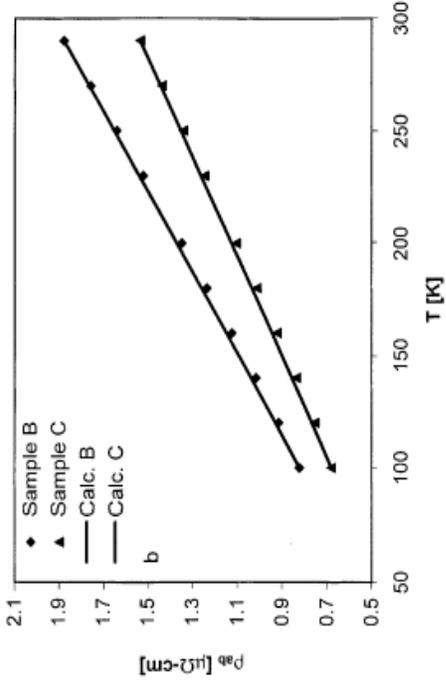
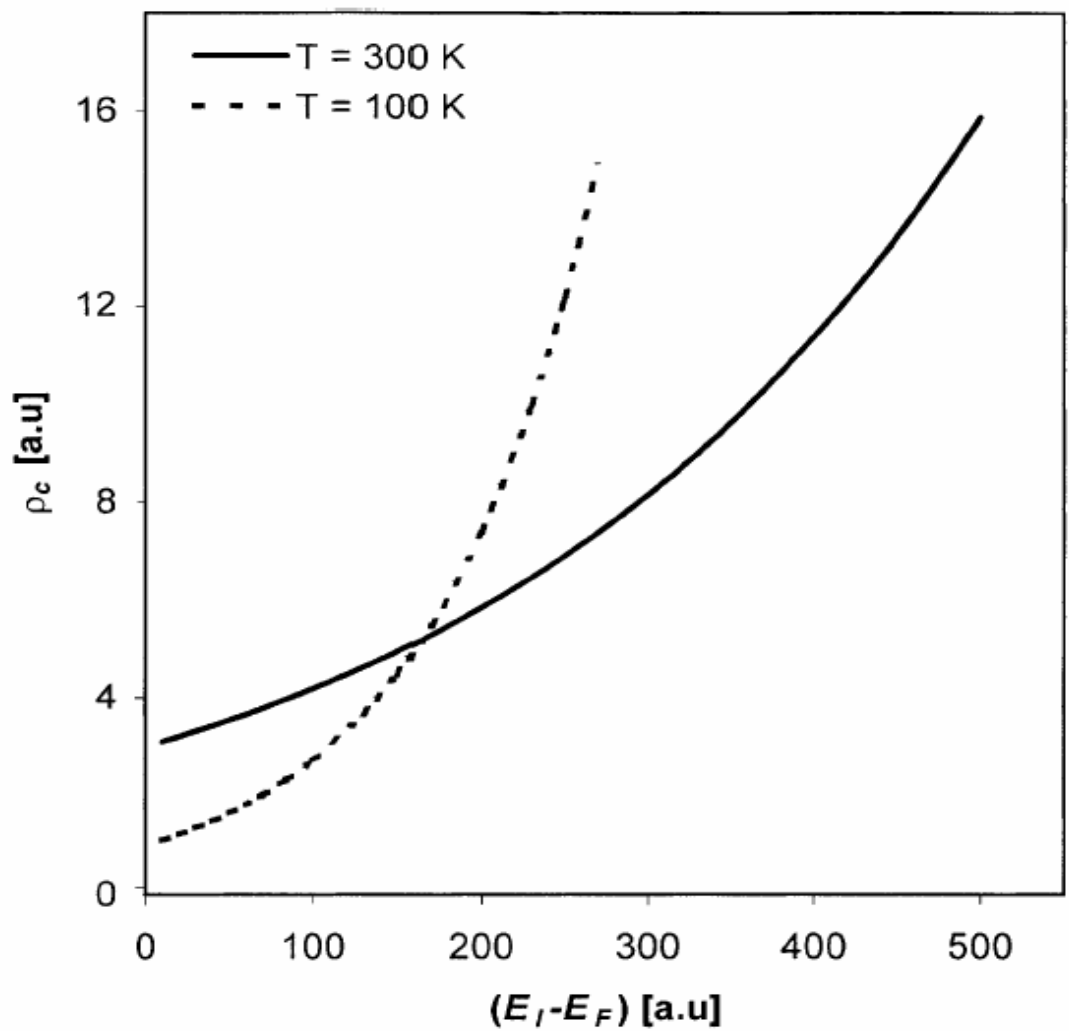
\end{document}